\documentstyle[12pt,epsf,a4wide]{article}
\begin{document} 

\newcommand{\ba}{\begin{eqnarray}}
\newcommand{\ea}{\end{eqnarray}}
\newcommand{\qt}{{q}}
\newcommand{\ut}{{u}}
\renewcommand{\kappa}{{k}}
\newcommand{\nd}[1]{/\hspace{-0.4em} #1}
\begin{titlepage}

\begin{centering}
\begin{flushright}
astro-ph/0108295 \\
CERN-TH/2001-226 \\
ACT-08-01 \\ 
CTP-TAMU-26-01 \\
\end{flushright}

\vspace{0.1in}

{\Large {\bf Non-Critical Liouville String Escapes Constraints on
Generic Models of Quantum Gravity}}

\vspace{0.4in}

{\bf John Ellis$^a$, N. E. Mavromatos$^b$} and {\bf D.V. Nanopoulos$^c$ } 
\\
\vspace{0.2in}

\vspace{0.4in}
 {\bf Abstract}

\end{centering}

\vspace{0.2in}

{\small It has recently been pointed out that generic models of quantum
gravity must contend with severe phenomenological constraints imposed by
gravitational \v{C}erenkov radiation, neutrino oscillations and the cosmic
microwave background radiation. We show how the non-critical
Liouville-string model of quantum gravity we have proposed escapes these
constraints. It gives energetic particles {\it subluminal} velocities,
obviating the danger of gravitational \v{C}erenkov radiation.  The effect
on neutrino propagation is naturally {\it flavour-independent}, obviating
any impact on oscillation phenomenology. Deviations from the expected
black-body spectrum and the effects of time delays and stochastic
fluctuations in the propagation of cosmic microwave background photons are
negligible, as are their effects on observable spectral lines from
high-redshift astrophysical objects.}

\vspace{0.8in}
\begin{flushleft}
CERN-TH/2001-226 \\
August 2001
\end{flushleft}

\vspace{0.5in}
\begin{flushleft} 

$^a$ Theoretical Physics Division, CERN, CH-1211 Geneva 23, Switzerland
\\
$^b$ Theoretical Physics Group, Department of Physics, 
King's College London, Strand, London WC2R 2LS, U.K. \\
$^c$ Department of Physics, Texas A \& M University, 
College Station, TX~77843, USA;
Astroparticle Physics Group, Houston
Advanced Research Center (HARC), 
Mitchell Campus,
Woodlands, TX~77381, USA;
Chair of Theoretical Physics, 
Academy of Athens, 
Division of Natural Sciences, 
28~Panepistimiou Avenue, 
Athens 10679, Greece. 
\end{flushleft}

\end{titlepage}

\newpage

\section{Introduction} 

A testable quantum `Theory of Everything', including gravity, is the
analogue for particle physicists of the `faint blue dot' towards which
much space research is directed. One might seek to test quantum theories
of gravity either through subtle effects in a conventional physical
framework, such as non-renormalizable operators scaled by some inverse
power of the Planck mass $m_P$, or through some qualitatively new effect,
such as a violation of quantum mechanics and/or relativity. The latter is
a rather dangerous route for models of quantum gravity to follow, since
both quantum mechanics and relativity are known to be very robust theories
that are difficult to make `slightly pregnant': any deviation from their
sacred canons risks catastrophic phenomenological consequences.

We have proposed a formulation of quantum gravity based on non-critical
string theory~\cite{ddk,aben} with a Liouville field playing the r\^{o}le
of time~\cite{emn}. This model suggests that there might be observable
violations of quantum mechanics, and that energetic particles might travel
at less than the velocity of (low-frequency) light. Liouville string
theory passes many non-trivial consistency tests, e.g., of the
interpretation of the Liouville field as time~\cite{emn}, the appearance
of an eleventh dimension in $M$ theory~\cite{emnM}, and the Helmholtz
conditions for quantization~\cite{emnI} in theory space.

Several accelerator and non-accelerator experiments are already sensitive
to violations of quantum mechanics and/or relativity that challenge
candidate models of quantum gravity. For example, probes of quantum
mechanics in the $K^0 - {\overline K^0}$
system~\cite{ehns,kaons}~\footnote{See also~\cite{bmeson} for a discussion
of similar effects in the $B^0 - {\overline B^0}$ system.} are sensitive
to violations of quantum mechanics that are scaled by one inverse power of
$m_P$~\cite{cplear}, and probes of light propagation from distant
astrophysical sources are sensitive to deviations from special relativity
that are also scaled by one inverse power of $m_P$. The latter might also
play a r\^{o}le in the possible evasion~\cite{gzkobs} of the
Greisen-Zatsepin-Kuzmin cutoff~\cite{gzkcutoff} by ultra-high-energy
cosmic rays (UHECR)~\cite{uhecr} and energetic astrophysical photons.

Recently, several potential pitfalls for candidate quantum theories of
gravity have been pointed out. In theories in which energetic particles
travel faster than low-energy gravitons, UHECR emit
gravitational \v{C}erenkov radiation at catastrophic rates, causing them
to lose energy and become unobservable~\cite{nelson}. Neutrino-oscillation
phenomenology would be drastically modified in candidate quantum theories
of gravity that have flavour-dependent effects on neutrino
propagation~\cite{brustein}. Most recently, it has been argued that the
light-cone fluctuations expected in generic quantum theories of gravity
have problematic consequences for the cosmic microwave background
radiation~\cite{distefano}. 

It is the point of the present note to demonstrate how non-critical
Liouville string escapes these potentially fatal constraints. 

In this approach, UHECR automatically travel {\it subluminally}, so the
issue of gravitational \v{C}erenkov radiation does not arise. The
quantum-gravitational effects on neutrino propagation are naturally {\it
flavour-independent}, so that effects on neutrino-oscillation
phenomenology are negligible. Finally, the effects on the black-body
spectrum of the cosmic microwave background radiation, including those due
to the expected delays and fluctuations in the propagation of microwave
background photons, are also negligible, as are their expected effects on
the spectral lines of astrophysical sources at high redshifts.

\section{Gravitational Cerenkov Radiation and Non-Critical String Theory }

We recall that an important constraint on some models of quantum gravity
is imposed by the absence of Gravitational Cerenkov radiation (GCR). This
was recently pointed out in~\cite{nelson}, in the context of a $D$-brane
scenario of warped gravity~\cite{rs}, but the remark is more general. 

In the warped-gravity framework, our four-dimensional world is given a
dual interpretation~\cite{4d} as a warped space-time~\cite{rs}, associated
with a $D3$-brane embedded in a higher-dimensional (bulk) geometry.  The
purely four-dimensional interpretation of such geometries is due to the
AdS/CFT correspondence~\cite{ads}. The ensuing violation of
four-dimensional Poincar\'{e} invariance implies that gravitational waves
on our brane world always travel with a speed $c_G$ {\it lower} than the
speed of light $c$. The fact that gravity propagates at a speed less than
that of light implies the appearance of GCR in such models~\cite{nelson},
in analogy with the conventional electromagnetic Cerenkov radiation (ECR)
emitted by any charged particle propagating in a medium where the velocity
of light is less than that of the particle, which may occur when the
medium's refractive index $n > 1$. ECR is emitted because the charged
particle outruns its own electromagnetic field in the medium, propagating
`superluminally'.  In a similar spirit, if gravitational waves propagate
slower than light or ultrarelativistic particles in our brane world, GCR
pheomenon will occur, even if the particle is uncharged, since all
particles couple to gravity.

Based on this observation, the observed UHECR were used to derive
restrictions~\cite{nelson} on the deviation of the `gravitational
refractive index' $n_G$ from 1, placing a lower bound on the propagation
speed of gravity in such warped space-time scenarios.  Simple kinematics
fix the rate of loss of energy due to GCR, which may be used to estimate
the maximum total time of travel possible for the observed
UHECR~\cite{nelson}: 
\begin{equation}
t_{\rm max} = M_P^2/(n_G - 1)^2 p^3
\label{gcrtime}
\end{equation}
where $M_P$ is the Planck mass, and $n_G >1$ is the gravitational
refractive index.  The observed UHECR with energies higher that $10^{11}$
GeV require $n_G - 1 < 10^{-15}$, and an even more stringent bound:  $n_G
- 1 < 10^{-19}$ can be derived in some models for the production of UHECR. 
These limits may be problematic for some warped-space-time models.

This argument can also be used to constrain models in which low-energy
gravitons, photons and other particles travel at speeds close to the
velocity of light, so that effectively $n, n_G \approx 1$ at low
frequencies, whereas ultra-high-energy particles travel faster than the
speed of light:  $n_{UHE} < 1$. Specifically, an analysis similar to that
of~\cite{nelson} shows that one also has
\begin{equation}
t_{\rm max} = M_P^2/(1 - n_{UHE})^2 p^3
\label{gcrtimeUHE}
\end{equation} 
An example of such a model would be one in which energetic particles
travel at velocities that differ from the speed of light by amounts that
increase with energy:
\begin{equation}
v(E) = c \left( 1 + \xi \left( \frac{E}{M_{QG}} \right)^p + \cdots
\right),
\label{badnews}
\end{equation}
with~\footnote{We may always absorb the magnitude of $\xi$
into that of the quantum-gravity scale $m_{QG}$, so that the two
distinguishable cases are $\xi = \pm 1$.}  $\xi = + 1$
and $p > 0$, resulting in
$1 - n_{UHE} \simeq (E / m_{QG})^p$. In this case, the bound
of~\cite{nelson} may be translated into
\begin{equation}
m_{QG} > 10^{11 + 15/p}~{\rm GeV},
\label{GCRboundonmQG}
\end{equation}
which might be problematic for models with $m_{QG} \sim 10^{19}$~GeV and
$p = 1$, in particular.

In our non-critical Liouville model for quantum gravity, we have predicted
that the velocities of energetic particles should deviate from the speed
of light in a manner similar to (\ref{badnews}) with $p = 1$. {\it
However}, in our model, we have also predicted that $\xi = - 1$, i.e.,
{\it energetic particles travel at less than the speed of light}.  Hence
the constraint (\ref{GCRboundonmQG}) {\it does not apply}.  We now give an
intuitive explanation of the sign of $\xi$, and then recall the formal
derivation that $\xi = -1$.

We envisage that, in any quantum theory of gravity, space-time will no
longer be (approximately) flat at short distances, since quantum
fluctuations in the background metric will endow it with a `foamy'
structure. This will give rise to microscopic `gravitational lensing'
phenomena, in particular time delays compared to propagation in a flat
space-time. These we expect to be relatively unimportant for
long-wavelength probes that average out many microscopic fluctuations,
whereas very-short-wavelength probes would be systematically delayed, as
they `bump over' all the corrugations in space-time. This rough and
intuitive picture suggests that $\xi$ should be negative and $p > 0$, as
we find in detailed calculations in our non-critical Liouville model
for quantum gravity~\cite{recoil} that we now review. 

In this approach, one may view our world as a three-brane punctured by
$D$-particle defects that deform space-time in their neighbourhoods when
struck by an energetic particle/string state. The presence of recoiling
defects in our brane world breaks explicitly Poincar\'{e} invariance, and
Lorentz invariance in particular. We have dealt specifically with scalar
closed-string modes for illustrative purposes. However, the results can
easily be extended to particles of non-zero spin. In particular, we have
derived~\cite{BI,szabo} 
a Born-Infeld modification of Maxwell electrodynamics, and shown
that photons propagate at {\it less} than the speed of light in the
corresponding modifications of Maxwell's equations. We have also
demonstrated a corresponding modification of the speeds of energetic
fermions~\cite{volkov}, to which we return anon. 

The basic idea is that, when an energetic string state strikes a
$D$-particle defect in our four-dimensional brane world, the latter
recoils at the moment of impulse, and hence the space-time surrounding the
defect is distorted.  Formally, the distortion of space-time is described
by a deviation from the conformal symmetry of the string theory
background, which is restored by Liouville dressing, with the Liouville
mode identified as target time~\cite{emn}.  With such a dressing, one
obtains, in the mean-field approximation, a distortion of the initially
flat space-time characterized by
the following non-zero off-diagonal element in the metric tensor: 
\begin{equation}
G_{0i}=u_i \Theta (t), \qquad t  \gg 1,
\label{recoilmetric}
\end{equation}
where $u_i \sim E / m_D$ is the recoil velocity of the $D$ particle of
mass $M_D \sim M_s / g_s: M_s=1/\ell_s$ is the string scale, and $g_s$
is the string coupling. The corresponding modification of
Maxwell's equations and the ensuing reduction in the velocity of light is
discussed in~\cite{Mitsou}.

The mean-field quantum-gravity result (\ref{recoilmetric})
leads, when one considers null geodesics in the background 
metric, to an {\it energy-dependent} refractive index $n_{QG}(E)$:
\begin{equation}
\label{refrindex} 
n_{QG}(E) = c\left(1 + \xi (\frac{E}{M_D}) \right)
\end{equation}
where $\xi = - |{\cal O}(1)| < 0$.
The appearance of the $D$-brane scale, replacing the generic
quantum-gravity scale $m_{QG}$, is a consequence of the
r\^ole of $D$-particle excitations as the dominant ingredients in
quantum space-time foam in this model. As mentioned above,
the appearance of a {\it subluminal} effective speed of light 
is a consequence of the Born-Infeld dynamics induced by the 
recoiling $D$ particles~\cite{szabo,recoil}.

In addition to this mean-field refractive-index effect, there are {\it
stochastic} effects on particle propagation. To understand their origin,
one must distinguish our foam model from a situation in which $D$-particle
defects are real. In our foam model, the defects are viewed as {\it
virtual} excitations/fluctuations of the quantum stringy gravity vacuum.
In a naive diagrammatic sense, therefore, one may consider a self-energy
loop diagram for a propagating particle, in which a $D$-particle defect is
emitted and then re-absorbed after a time interval $t_D^{\rm stoch} \sim
t_P$.  The emission process causes the particle to `lose' energy to the
defect, which is `regained' after a time $\sim t_P$ during the
reabsorption. Thus, in space-time foam energy is conserved on {\it the
average} for the propagation of particles~\footnote{However, energy may
not be conserved in decays~\cite{GZKemn}.}.
During the emission and re-absorption
processes, there is a non-trivial gravitational distortion of space-time,
which results in the formation of a `bubble' with a refractive index that
fluctuates non-trivially, as discussed in~\cite{recoil} and reviewed
briefly here.

The interaction of an energetic particle with the $D$-particle defects 
leads, in good approximation, to Gaussian 
fluctuations in the recoil velocity of the $D$ particle,
with a width of order~\cite{recoil}: 
\begin{equation}\label{stochu} 
\Delta^{\rm stoch} u_i = \Gamma - \Gamma (u=0) = 
{\cal O}\left(g_s^2\frac{{\overline u}^2}{2}\right)
\end{equation}
These uncertainties in the recoil velocity `lost' to
the foam imply corresponding fluctuations in the energy $\Delta E$ of the 
propagating particle:
\begin{eqnarray}\label{stochu2}
\Delta^{\rm stoch} E \sim \Delta E_{\rm kin} \sim
|{\overline u}|^2 \frac{\Delta^{\rm stoch} u}{u}M_D
\sim \frac{g_s^2\, E^3}{M_D^2}
\end{eqnarray}
These fluctuations can be related to the resummation 
of world-sheet genera that
corresponds to higher-order quantum effects in the $D$-particle model of 
space-time foam. Formally, they arise because of modular 
infinities associated with specific deformations of the 
world-sheet $\sigma$ model
that have zero conformal dimension, and are related to the logarithmic 
operators characterizing the $D$-particle recoil~\cite{recoil}. 

\begin{figure}
\epsfxsize=3.5in
\bigskip
\centerline{\epsffile{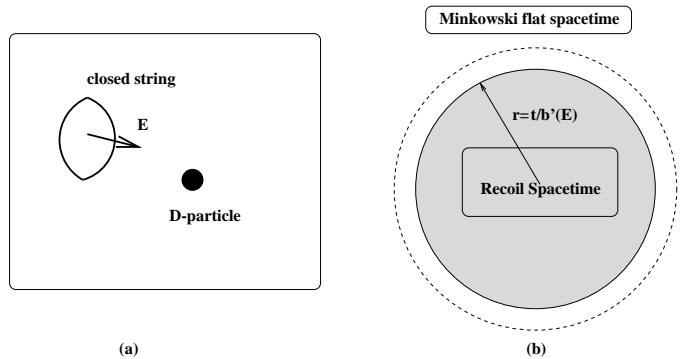}}
\vspace{0.2in}
\caption{\it (a) Scattering of a closed-string mode
of energy $E$ off a $D$ particle
embedded in a four-dimensional space-time. (b) After the recoil
an effective bubble (shaded region) is formed, 
inside which particles travel at less than
the (low-energy) speed of light, whereas the  exterior geometry is
flat Minkowski. The bubble is unstable and 
shrinks by evaporation, as indicated by the dashed circle.}
\label{fig:bubble}
\end{figure}

One may treat such effects by
regarding~\cite{recoil} the disturbance of space-time due to the recoil of
the heavy defect as creating a microscopic space-time `bubble', as seen in
Fig.~\ref{fig:bubble}, whose radius is of the same order as the inverse
quantum uncertainty $b'(E)$ in the momentum of the heavy $D$
particle~\cite{szabo}:
\begin{equation} 
{b'}(E)^2 = 4g_s^2\,M_s \left(1-\frac{285}{18}g_s^2\frac{E^2}{M_D^2} + 
\dots \right)
\label{uncert}
\end{equation} 
where $E$ again denotes the energy of the incident particle. We note that
$b'(E)$ decreases with increasing energy~\cite{szabo}.
As argued in~\cite{recoil}, the formation of such `bubbles' induces
quantum fluctuations in the two-point function for 
the metric (\ref{recoilmetric}),
which in turn induce stochastic corrections in
$G_{00}$ component of the metric:  
\begin{equation} 
\Delta G_{00} =\frac{{b'(E)}^2 r^2}{t^2}\, : \qquad r^2 =
\sum_{i=1}^{3} x_i^2,
\label{stochasticmetric} 
\end{equation} 
where $b'(E)$ is given by (\ref{uncert}) when $E \ll M_D$.

\begin{figure}
\epsfxsize=3.5in
\bigskip
\centerline{\epsffile{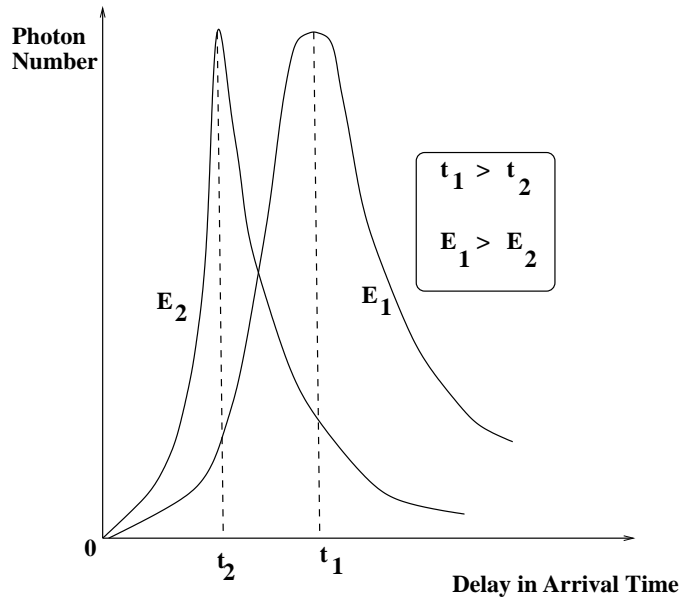}}
\vspace{0.2in}
\caption{\it Illustration of the quantum-gravity 
induced refractive index (\ref{refrindex}) and 
stochastic effect (\ref{stochindex}).
The delays in the arrival times $t_i$ of photons increase with 
increasing energy $E_i$, due to the subluminal character of 
the effect (\ref{refrindex}). The stochastic effects (\ref{stochindex}),
on the other hand, imply that the width of a
pulse of photons with the same energy spreads out more 
for the more energetic channels, although the magnitude of the effect is 
smaller than the retardation.}  
\label{fig:delays}
\end{figure}

The energy-independent part of (\ref{stochasticmetric}) proportional to
the first term in (\ref{uncert}), i.e., proportional to $g_s^2$, should
be subtracted and absorbed into the conventional Minkowski metric. The
remaining $E$-dependent parts in (\ref{stochasticmetric}), when computed
near the light-cone $r \sim t/2g_s$ where the recoil space-time is
matched with the external Minkowski space-time~\cite{gravanis}, yield
stochastic fluctuations about the previous mean-field refractive-index
effect (\ref{refrindex}): 
\begin{equation}
\left(\Delta n_{QG}(E)\right)_{\rm stoch} = 
{\cal O}\left(g_s^4\frac{E^2}{M_D^2}\right)
\label{stochindex}
\end{equation}  
We observe that this stochastic effect is suppressed relative to the
mean-field refractive-index effect (\ref{refrindex}) by two powers of 
$g_s^2$ and of $E / M_D$~\footnote{In our previous phenomenological 
analysis~\cite{Mitsou}, we considered also suppression of this effect by 
a single power of $E / M_D$, but (\ref{stochindex}) has a better 
justification within our approach.}. Hence, the dominant effect in this 
approach will 
be a reduction in the resulting speed of energetic particles, with small
fluctuations around the mean-field velocity~\footnote{For later
reference, we recall that the above phenomenon may also be interpreted as a 
fluctuation in the light cone, given the refractive 
index mean field effect (\ref{refrindex}).}, as seen in 
Fig.~\ref{fig:delays}.

We conclude that particle propagation is always {\it subluminal} in this
approach, a property traceable to the characteristic Born-Infeld dynamics
underlying the $D$-particle recoil process~\cite{recoil}. Thus, in
contrast to the case considered by~\cite{nelson}, energetic particles
propagate with increasingly subluminal velocities, and hence no GCR or ECR
is emitted. 

The non-trivial refractive index (\ref{refrindex}) has potentially
observable effects on the arrival times of energetic particles from
distant astrophysical objects such as gamma-ray bursters~\cite{nature}.
These and other suitable sources have been studied~\cite{Mitsou,others}, 
with the conclusion
that $m_{QG} > 10^{15}$~GeV. Some tests of this retardation effect have
sensitivity to larger values of $m_{QG}$, 
but the corresponding bounds
cannot yet be regarded as statistically established. 

\section{Neutrino Oscillations in Non-Critical String Theory}

We now come to discuss a second constraint~\cite{brustein}
on generic quantum gravity models, provided by the phenomenology of
neutrino oscillations. It has been observed that, 
if the modifications to neutrino energy-momentum dispersion
relations induced by quantum gravity are {\it flavour-dependent}:
$M_{QG} \to M_{{QG}_i}$, oscillation patterns will be affected on a
distance scale $L$ of order:
\ba
L_{ij} \times \left( {1 \over M_{{QG}_i}} - {1 \over M_{{QG}_j}} \right) 
\sim 2\pi \frac{1}{E^2}
\ea
where $E$ is the typical energy.
The point of \cite{brustein} is that, if $M_{{QG}_i} - M_{{QG}_j} \sim M_P
\sim 10^{19}$~GeV, the induced $L_{ij} \sim 10$ Km for $E \sim 1$~GeV,
which is much smaller than the characteristic oscillation length of
atmospheric neutrinos, which have energies $\sim 1$~GeV. It is concluded
that if the $M_{{QG}_i} \sim 10^{19}$~GeV they must be very nearly equal.
Alternatively, the $M_{{QG}_i} \gg 10^{19}$~GeV.

However, as was mentioned already in~\cite{volkov}, in the context of a
general discussion of fermion propagation through space-time foam in our
non-critical Liouville string model, we expect the recoil and bubble
formation effects discussed in the previous section to be essentially
kinematical and geometrical, depending only on the energy of the
propagating particle.  As such, these phenomena should exhibit an
extension of the equivalence principle, and hence be {\it
flavour-independent}, with no effects on oscillations~\footnote{We also
recall that decoherence effects on neutrino 
propagation~\cite{lisi,adler} are
also not expected~\cite{emnadler} to have observable effects on present
neutrino-oscillation experiments.}.

We recall that, since ultra-high-energy neutrinos are not absorbed by
known cosmological backgrounds, they are not subject to a GZK cutoff
analogous to that expected for ultra-high-energy protons and energetic
photons. This provides the opportunity for probing particle propagation
through the Universe over long distances at unequalled energies. For
example, if gamma-ray bursters at cosmological distances $\sim 3000$~Mpc
emit a detectable flux of neutrinos at energies up to $10^{10}$~GeV, as
suggested in some astrophysical models, it will be possible to test 
quantum-gravity models with $M_{QG}
\sim 10^{27}$~GeV. Alternatively, if $M_{{QG}} \sim 10^{19}$~GeV, such
ultra-high-energy neutrino bursts would be spread out over many years, and
hence unobservable~\cite{volkov}. 

\section{Possible Signatures in the Cosmic Microwave Background Radiation} 

Another potential cosmological pitfall for quantum theories of gravity has
recently been pointed out~\cite{distefano}, namely a possible observable
distortion of the cosmic microwave background (CMB) radiation spectrum. 
That pioneering analysis was done in the context of a model of light-cone
fluctuations. Here we treat the question from the point of view
of~\cite{recoil}. We examine here the possible effects at the source - does
the thermal spectrum in a quantum-gravity model differ from that in
conventional thermodynamics - and during propagation from the last
scattering surface - are there observable effects of the non-trivial
refractive index and stochastic effects? 

The possible effect on the thermal spectrum of the CMB has to be
considered from the point of view of our identification of the Liouville
field with time. At finite temperatures, following the Matubara
formulation of thermal field theory, we Euclideanize the Minkowski time by
analytic continuation, and then compactify the time on a circle of radius
$\beta$, the inverse temperature. We observe that, by construction,
finite-temperature effects induce non-criticality in string theory,
because the associated excitation spectrum of the finite-temperature
string is in one dimension lower. Even if one started from a string theory
in the critical dimension, the resulting finite-temperature string lives
in a non-critical dimension. As a result, the induced central-charge
deficit $Q$ always contains a term proportional to the physical
temperature. In our approach of identifying time with the Liouville mode,
additional departures from criticality (before compactification of the
time coordinate) will also contribute to the induced central-charge
deficit of the Liouville theory, and thus also to the temperature. This
should be understood as a sort of back-reaction effect.

In the context of non-critical Liouville string,
this process has non-trivial consequences, which we now 
analyze~\cite{recoil}. 
First, we remark that, 
in Liouville strings, the dilaton field $\Phi$ contains a term linear in
the Liouville field $\phi$:
\begin{equation} 
\Phi \ni \frac{1}{2}Q\phi + \dots
\end{equation}
where $Q$ is the central-charge deficit.
Therefore, when we Euclideanize the Liouville time variable $\phi$,
the $\sigma$-model dilaton term, which couples to the world-sheet 
curvature $R^{(2)}$,  becomes~\footnote{We concentrate on the 
world-sheet zero mode of the Liouville field.}:
\begin{equation} 
\frac{1}{2\pi}\int _{\Sigma} \Phi R^{(2)} \to  
i\chi Q\phi_E   
\end{equation}
where $\chi$ is the Euler characteristic, which is $\chi=2$ for a
world sheet with spherical topology.
If one makes $\phi_E$ periodic with period $2\pi\beta$, then,
in our normalization of the Liouville terms:
\begin{equation}
  2\beta Q = 1 
\end{equation}
i.e., the Liouville temperature is determined by the 
central-charge deficit $Q$. 

In non-critical string, the central-charge deficit $Q$  
is in general a functional of the couplings of the various 
deformations of the $\sigma$ model that define the theory away from 
its conformal point. 
In the specific $D$-brane model~\cite{recoil}, such couplings are 
essentially the recoil velocities $u_D$ of the $D$ particles in the foam.
The corresponding deficit is given by $Q^2 = c[u_D]-c^*$,
where $c[u_D]$ is the `running central charge' of the 
deformed theory, and 
$c^*$ is 25 for bosonic strings or 10 for superstrings. The central-charge
deficit satisfies a Zamolodchikov flow equation:
\begin{equation} 
\frac{\partial Q^2}{\partial \tau} = - \beta^{u_D}{\cal G}_{DD}\beta^{u_D}
\end{equation} 
where $\tau$ is a renormalization-group scale on the world-sheet, 
$\beta^{u_D} = -\frac{\epsilon^2}{2}u_D + {\cal O}(u^2_D)$ 
is the world-sheet renormalization group $\beta$ function of the 
recoil operator,
and $\epsilon^{-2} \sim \tau$~\cite{recoil}. 
Since the Zamolodchikov metric ${\cal G}_{DD} \sim \frac{1}{\epsilon^2}$,  
and $u_D \sim \epsilon {\hat u}$, where ${\hat u}$ is independent
of the scale $\epsilon$~\cite{recoil}, 
the above flow equation may be integrated in the neighborhood of 
a fixed point, i.e., dropping terms of order higher than $u_D^2$,
to yield the total deficit:
\begin{equation}
Q \sim Q_0\sqrt{1 + {\cal O}(u_D^2)}.
\end{equation}
Here $Q_0$ a constant, associated with the equilibrium temperature
in the limit of infinitely heavy $D$ particles, where recoil (back-reaction)
effects are absent. The non-critical fixed point value $Q_0 \ne 0$,
because of the dimensional reduction in the spectrum of the corresponding 
string theory at finite temperatures that was mentioned above.

Energy-momentum conservation, which can be proven rigorously 
in our model~\cite{recoil},
implies that $u_D^2 \sim \left(g_s\frac{\Delta E}{M_s}\right)^2$, 
where $\Delta E$ is 
a typical energy transfer during the scattering of a matter particle on
a $D$ particle in the foam, i.e., $\Delta E ={\cal
O}(E)$, where $E$ is the energy of the incident particle.
In the case of photons with frequencies
$\nu$, including the CMB photons, 
the above analysis implies that there will be 
variations in the effective temperature, related to the photon
frequencies
and suppressed by a power of the Planck scale $M_s/g_s$. To leading order:
\begin{equation}\label{tfluc}
\delta \beta \sim {\cal O}\left(g_s^2 \frac{\nu^2}{M_s^2}\beta\right) =
{\cal O}\left(\frac{\nu^2}{M_P^2}\beta\right).
\end{equation}
We consider now the effect on a thermal spectrum of frequencies $\nu$,
looking for deviations from the Boltzmann form:
\begin{equation}\label{thermal1} 
F_0(\nu)=\frac{a\, \nu ^3}{e^{\beta \nu}-1}
\end{equation}
where $a$, $\beta$ are constants.  The fluctuations described above
(\ref{tfluc}) cause deviations from the naive thermal spectrum of order: 
\begin{equation} 
\delta F_0 \simeq 
F_0 \frac{\nu \delta \beta \, e^{\beta \nu}}{e^{\beta \nu}-1}=
F_0 \frac{\nu^3\beta}{M_P^2}\frac{e^{\beta \nu}}{e^{\beta \nu}-1} =
{\cal O}\left(10^{-56}\, F_0\right),
\end{equation} 
for $T \sim 3000$~K and $\nu \simeq 1$~eV, typical values when the CMB
was produced. This is undetectable in practice. 

We now turn to the question whether effects on the propagation of 
CMB photons since their last scattering surface could have observable 
effects. As discussed in~\cite{Mitsou}, the difference in travel times 
(distances travelled) of 
two photons with present energies $E_{1,2}$ following emission at 
redshift $z$ is
\begin{equation}
\Delta t = {\Delta L \over c} = {2 \over H_0} \left[ 1 - {1 \over \left( 1 
+ z \right)^{1/2}} \right] { E_1 - E_2 \over M_{QG}}.
\label{mitsou}
\end{equation}
For $E_1 - E_2 \sim 10^{-3}$~eV and $m_{QG} \sim 10^{19}$~GeV, we find
$\Delta L \sim 10^{-3}$~cm, which is negligible compared with the 
thickness of the last scattering surface of the CMB. We conclude that the 
mean-field refractive-index effect is also undetectable in the CMB. Since 
the stochastic effects (\ref{stochindex}) on photon propagation are even 
smaller, the same conclusion applies to them, in our model.

As already commented, stochastic fluctuations in the photon velocity have
effects equivalent to light-cone fluctuations. It has been
suggested~\cite{distefano} that these might be observable in the CMB data
in models~\cite{ford} involving large compactified extra dimensions. We
now contrast the situation in such models with the stochastic light-cone
fluctuations expected in our model of Liouville quantum
gravity~\cite{recoil}, which cannot be constrained by CMB data.

We consider a thermal spectrum (\ref{thermal1}) 
of frequencies $\nu$, as appropriate for a CMB background. 
In this case, with a temperature $T=2.725$ K, the constant $\beta$ in 
(\ref{thermal1}) may be expressed as $\beta = 1.76 \times 10^{-11}$ s. 
In our model, a monochromatic wave of frequency
$\nu_0$ becomes a Gaussian with width:
\begin{equation} \label{dvu}
      \Delta \nu|_{\rm Bubble} = \left({\nu_0^3 \over M_D^2} \right)
\end{equation}
each time a `bubble' is formed. Expecting these to have a characteristic 
size $\sim \ell_P$, the total Gaussian width we expect is:
\begin{equation} \label{realdvu}  
      \Delta \nu|_{\rm Recoil} = \left({\nu_0^3 \over M_D^2} 
\right) \sqrt{L \over \ell_P}.
\end{equation}
In the model of~\cite{distefano}, there is a characteristic
light-cone fluctuation time $\Delta t$ which induces:
\begin{equation} 
\Delta \nu |_{\rm Light-cone}= \nu_0^2 \Delta t\, , 
\label{fordnu}
\end{equation} 
Using the available CMB data, it has been argued~\cite{distefano} 
that $\Delta t < 2.1 \times 10^{-14}~{\rm s}$.
In our case, we expect that in (\ref{realdvu}) one should use
$\nu_0 \sim 10^{-3}$~eV, $M_D \sim 10^{19}$~GeV, $L \sim 10^{10}$~ly and 
$\ell_P \sim 10^{-33}$~cm, corresponding in the notation 
of~\cite{distefano} to $\Delta t \sim (\nu_0 / M_D^2) \sqrt{L \over
\ell_P} \sim 10^{-43}$~s. This confirms our previous estimate that
effects on the propagation of CMB photons are negligible in our 
non-critical Liouville-string model of space-time foam.

One may also consider stochastic (or light-cone fluctuation) effects on 
lines observable in the spectra of distant astrophysical objects with $z = 
{\cal O}(1)$. As in (\ref{realdvu}), we expect such spectral lines to be 
broadened by an fraction:
\begin{equation} 
\frac{\Delta \nu}{\nu} 
\sim \left( \frac{\nu^2}{M_D^2} \right) \sqrt{\frac{L}{\ell_P}}
\label{line}
\end{equation}
For distances at $z \sim 1$ one has 
$L/\ell_P \sim 10^{61}$, and one has $\nu/M_D \sim 
10^{-28}$ for typical atomic lines, implying
$\Delta \nu /\nu \sim 10^{-26}$, which is again undetectable. 

\section{Conclusions}

We have explored further in the present article a model for space-time
foam based on a $D$-brane world punctured by fluctuating virtual
$D$-particle defects. We have explained why certain phenomena that can
severely constrain other quantum-gravity models do not apply to ours.

Gravitational \v{C}erenkov radiation is absent in our model, because
energetic particles such as UHECR travel {\it subluminally}. This has been 
demonstrated formally, but the intuition is simply that more-energetic 
(higher-frequency, shorter-wavelength) particles feel more strongly the 
iregularities (bumps) in the space-time foam.

Neutrino oscillations are absent in our model, because the effects on
energetic neutrino propagation that we find are {\it flavour-independent},
and so do not affect oscillation phenomenology. The basic reason for this
is that the violations of four-dimensional Poincar\'{e} and Lorentz
invariance that we find are {\it purely kinematic and geometric} in
origin.

We have estimated the effects on the CMB of expected deviations from the 
initial thermal spectrum, and those of our mean-field refractive-index 
effect and stochastic light-cone fluctuations. In each case, we find 
effects considerably smaller than the observational upper limits. The same 
is true for the spreading of spectral lines predicted in our approach.

We conclude that, although our model for space-time foam is quite radical, 
predicting deviations from conventional quantum mechanics and 
relativity that are prospectively disastrous, it escapes like Houdini from 
the constraints considered in this paper.

\section*{Acknowledgements}

N.E.M. wishes to thank H. Hofer (ETH, Zurich and CERN)
for his interest and partial support.
The work of D.V.N. is supported by D.O.E. grant 
DE-F-G03-95-ER-40917.

\end{document}